\documentclass{aa}
\usepackage{amssymb}
\usepackage{amsmath}
\usepackage{txfonts}
\usepackage{graphicx,times,url,xspace}
\usepackage{natbib}
\usepackage{rotating}
\usepackage{url}
\usepackage{epsfig}
\usepackage{longtable}
\def\fermi{{\it Fermi}\xspace}

\begin{document}

\title{Cold ultrarelativistic pulsar winds as potential sources  of  galactic  gamma-ray lines above 100~GeV}
\author{
  Felix  Aharonian \inst{1,2}, Dmitry  Khangulyan \inst{3}, Denis Malyshev\inst{4}
}

\institute{
  Dublin Institute for Advanced Studies, 31 Fitzwiliam Place, Dublin 2, Ireland \\
  \and
  Max-Planck-Institut fur Kernphysik, Postfach 103980, 69029 Heidelberg, Germany \\
  \and
  Institute of Space and Astronautical Science/JAXA, 3-1-1 Yoshinodai, Chuo-ku, Sagamihara, Kanagawa 252-5210, Japan \\
  \and
  Bogolyubov Institute for Theoretical Physics, Metrologichna str., 14-b, Kiev 03680, Ukraine
}

\authorrunning{Aharonian et al.}
\titlerunning{Klein-Nishina gamma lines in the Galactic spectrum}

\abstract
{The  evidence of a   line-like   spectral features above 100~GeV, in particular at  130~GeV, recently reported from some parts of the galactic plane poses serious  
challenges  for any interpretation of this surprise discovery. It is generally 
believed that  the  unusually  narrow profile of the spectral  line
cannot be explained by conventional processes in  astrophysical objects,  and, if real,  is 
likely to  be associated with Dark Matter.}
{In this paper we argue that cold ultrarelativistic pulsar winds can 
be alternative sources of  very narrow  gamma-ray lines.} 
{We demonstrate that Comptonization
of a cold ultrarelativistic electron-positron pulsar wind   in the deep Klein-Nishina regime can readily 
provide very narrow  ($\Delta E /E \leq 0.2$)  distinct gamma-ray line features.  
To verify  this prediction,  we produced  photon count maps based on the  Fermi  LAT data 
in the energy interval 100 to 140~GeV.}
{We confirm earlier reports of the  presence of marginal gamma-ray line-like 
signals from three regions of the galactic plane.  Although the maps show some structure inside 
these regions, unfortunately the limited photon statistics do not allow any firm conclusion in this regard.}
{The confirmation of  130 GeV  line emission  by low-energy threshold atmospheric Cherenkov telescope systems, in particular by the new 27~m diameter dish of the H.E.S.S.  array,  would be crucial for resolving the spatial structure of the reported  hotspots, and thus for distinguishing between the Dark Matter and Pulsar origins of the  `Fermi Lines'.   }

\maketitle

\section{Introduction}
Recent reports on the possible  presence  of a narrow line-like feature in the spectrum of gamma-ray emission at
130 GeV from the galactic center region, and, presumably,  from some other parts of the galactic
plane \citep{brigmann12,weniger12,tempel12,boyarsky12,su12}, have received a prompt and 
enthusiastic reaction from the astrophysics and
particle physics communities.  Despite the marginal statistical significance of the reported signals and some
outstanding questions and inconsistencies, the implications of these results are hotly debated, basically in the context
of Dark Matter (DM).  This is motivated not only by the recognition of the potential of gamma-ray observations for
indirect searches of DM \citep[for a recent review see][]{Bergstrom12} and the overall excitement caused by the possible
association of the gamma-ray line at 130~GeV with DM \citep[see e.g.][]{Buckley12}, but also by unusual characteristics
of the signal. It is generally believed that the width of the 130~GeV line (of a few tens of GeV) is too narrow
to be explained by any physical process except for annihilation of DM\footnote{Based on the analysis of \fermi LAT data
  by \citet{weniger12}, it has been argued \citep{Profumo12} that the contamination of a bright featureless power-law
  background component (related e.g. to the radiation of the interstellar medium) by hard photons arriving from Fermi
  bubbles with a sharp spectral break between 100 and 150~GeV, can mimic a spurious line.  However, a closer look at the
  morphology shows that the 130 GeV feature is  most likely not  
  associated with Fermi Bubbles \citep[see e.g.][]{tempel12}.}.  On
the other hand,  some other features, in particular the significant shift of the center of  gravity 
of the signal from the
position of the galactic center \citep{su12}, as well as the possible presence of gamma-ray lines at other energies
\citep{boyarsky12} and from other parts of the galactic plane \citep{tempel12,boyarsky12}, challenge the DM origin of the reported signal.
Unfortunately, the marginal gamma-ray photon statistics does not allow definite conclusions in this regard.  The situation will be somewhat improved after doubling the photon statistics above 100~GeV by observations of the galactic plane
with \fermi Large Area Telescope (LAT) over the next several years. The development of new dedicated approaches to the  data
reduction focused on the highest energy domain of \fermi LAT also may help to clarify some of the current uncertainties and inconsistencies. 
 Finally, there is  hope that soon the low-energy threshold imaging Cherenkov telescopes, in
particular the new 27m diameter dish of the H.E.S.S. telescope array located in the Southern Hemisphere (see
http://www.mpi-hd.mpg.de/hfm/HESS/), will greatly contribute to the clarification of the question concerning the origin
of $\geq 100$~GeV gamma-ray line(s) - are they {\it instrumental}, {\it cosmological (DM)}, or {\it astrophysical}?  

The last option implies production of gamma-rays by accelerated
particles. So far it  has been discarded, basically because of 
the common belief that conventional high energy processes with involvement of 
ultrarelativistic  particles could not  reproduce gamma-ray line features.  
In Sec.\ref{astrolines} we briefly discuss different  
gamma-ray production mechanisms in the VHE domain in the context of their ability to 
produce sharp gamma-ray lines, and argue that the 
inverse Compton scattering in the Klein-Nishina regime by cold
ultrarelativistic electron-positron pulsar winds can result in sharp gamma-ray line
emission. The conditions for
reproduction of narrow  Klein-Nishina line profiles 
 are discussed in Sec.\ref{KNlines}.  Finally, in Sec.\ref{Data} 
we describe our study  of the spatial distribution  of $E \geq 100$~GeV photons based on the {\it Fermi} LAT data.   We confirm the  results reported earlier  on the presence of marginal gamma-ray line-like 
signals from three regions of the galactic plane, but,  because of limited photon statistics, we cannot 
make a strong statement on   
the existence  of localized hot spots  inside  the  `Fermi Line'   regions . The  results and conclusions  are  summarized in Sec.\ref{summary}.  

\section{Astrophysical VHE gamma-ray lines?}
\label{astrolines}

Interactions of relativistic particles with a broad energy distribution   
cannot result in sharp gamma-ray spectral features.    Moreover, even in the case of interactions of monoenergetic relativistic particles, the energy
spectra  of resulting gamma-rays generally  are broad.  For example, a 1~TeV proton interacting 
with  surrounding  gas produces
gamma-rays (through production and decay of $\pi^0$- mesons) with an average energy of 100 GeV, but the distribution of
gamma-rays is very broad with $\Delta E/E \gg 1$ \citep[see e.g.][]{Kelner06}. Actually, this is true for all
hadronic interactions, including  photomeson processes, with involvement of  secondary 
$\pi^0$-mesons.  

Narrow  gamma-ray
distributions in principle can be expected from monoenergetic beams of ultrarelativistic heavy ions, e.g.  $^{56} \rm Fe$, excited at
interactions with surrounding low-frequency radiation.  The Doppler-boosted gamma-ray emission due to the
de-excitation of these nuclei with Lorentz factor $\Gamma \geq 10^5$ may lead, in principle, to a rather narrow lines at energy 
$E = \Gamma \ E^* \sim 100$~GeV (typically the prompt de-excitation gamma-ray lines are produced with energy
in the frame of the nucleus $E^* \sim 1$~MeV). However, the efficiency of this mechanism in typical astrophysical
environments is very low \citep{FATaylor10}.  Also, the disintegration of the primary nuclei would lead to emission of a
large number of lines from secondary nuclei, and eventually to a rather broad distribution.  

The synchrotron radiation of charged particles, electrons or  protons,  also leads to broad-band emission. 
Even in the case of radiation of  monoenergetic  particles in a homogeneous magnetic field,
the spectral energy distribution (SED) is broad,  $E^2 {\rm d} N/{\rm d} E\propto E^{4/3}
\exp{(-E/E_0)}$ with $\Delta E/E \geq 1$.
In addition to synchrotron radiation, there are two other processes for the 
production of high energy gamma-rays  by relativistic electrons - bremsstrahlung and inverse Compton scattering (IC).  Bremsstrahlung of
monoenergetic electrons results in very hard, but still continuous gamma-ray distribution, 
$E^2 {\rm d} N/{\rm d} E \propto E$.  One should mention also that at very high energies 
the electron cooling typically is dominated by  synchrotron and IC radiation losses,
thus  bremsstrahlung plays an important role
only at relatively low, sub-GeV energies.

In contrast, the inverse Compton scattering of relativistic electrons is a universal gamma-ray
production mechanism which can work with very high efficiency throughout the entire gamma-ray domain, from MeV to TeV
energies, in diverse astrophysical environments, from compact objects like pulsars and AGN  to extended sources like
supernova remnants and clusters of galaxies \citep[see e.g.][]{FA04}. In the Thompson regime, 
IC scattering boosts 
the energy of target photons to   $E_\gamma \propto \varepsilon \Gamma^2$, 
where $\varepsilon$ is the target photon energy, and
$\Gamma$ is the electron Lorentz factor.  Thus,  in order to produce a narrow gamma-ray spectral feature, one should require
very narrow distributions 
 for both relativistic electrons and target photons.  
However, even in this case,
the distribution of the upscattered radiation is relatively broad due to the character of the
cross-section in
the Thompson regime. Another principal limitation comes from the requirement of narrowness of the target photon
field. Even in the case of black-body radiation, the width of the photon distribution exceeds the average photon energy
$\varepsilon \approx 3 \ \rm kT$, thus the original distribution of target photons will be shifted to higher energies
by a factor of $\Gamma^2$, and non-negligibly broaden due to the cross-section in the Thompson regime.  Even in the case
of an abrupt cutoff in the electron spectrum and the Planckian distribution of target photons, the spectrum of
gamma-rays contains a rather long exponential tail \citep{Lefa12}.

The picture changes dramatically, however,  when the IC scattering proceeds in the Klein-Nishina regime, i.e.  when $b=\varepsilon
E_{\rm e}/m_{\rm e}^2 c^4 \geq 1$.  In this case, the target photons take the entire momentum of the electron, thus, 
independent of the energy distribution of target photons,  one may expect a line shape for  the upscattered
radiation if the electrons have a very narrow distribution.  Both conditions can be realized in pulsars, namely when the
cold ultrarelativistic electron-positron pulsar winds upscatter the surrounding high energy gamma-ray emission
\citep{Bogovalov00}.  For isolated pulsars, a sharp gamma-ray line is  produced at  Comptonization of the
electron-positron wind by the thermal X-ray emission from the surface of the neutron star. The mechanism can be
effective only if the conversion of the Poynting flux to kinetic energy of the bulk motion (``acceleration'' of the
wind) takes place close to the light cylinder \citep{Bogovalov00}. In the case of pulsars with bright nonthermal
broad-band magnetospheric emission, e.g. in the Crab pulsar, the major fraction of energy of the wind is channeled into
the pulsed VHE continuum. It is likely \citep{ABK_Nature12} that this component of radiation of the pulsar wind is
responsible for the recently detected VHE pulsed  radiation of the Crab pulsar \citep{VHEpulsed_VERITAS,VHEpulsed_MAGIC}.

Gamma-ray line-like structures are expected also from binary systems containing a pulsar and a luminous optical star \citep{wind_BallKirk00,Khangulyan07}. In
this case the efficiency of IC scattering is higher because of presence of copious target photons provided by the
companion star, but the gamma-ray lines are less sharp compared to the isolated pulsars because of 
smaller values of
the Klein-Nishina parameter $b$. 
It has been recently argued \citep{KABR12} that the bright flare of the 
binary pulsar PSR 1259-63/LS2883, detected by \fermi in
2011 \citep{1269flare_fermi,1259flare_tam}, could be best explained by the IC scattering of the unshocked pulsar.  The
spectral measurements of \fermi LAT require a relatively modest Lorentz factor of the wind, $\Gamma \approx 10^4$;
the IC scattering proceeds in the Thompson regime, therefore the resulting gamma-radiation has a rather broad energy
distribution \citep{KABR12}.

Finally, one should mention two other  (highly speculative)  scenarios of formation of a very narrow  multi-GeV spectral gamma-ray line feature. It  could be related to the  0.511~MeV line  produced at annihilation 
of ($e^+e^-$)  pairs in  cold plasma of  an outflow (jet or a wind) relativistically moving with Lorentz factor 
$\Gamma \geq  10^4$.  A very sharp increase of the gamma-ray spectrum can be expected also in a quite different scenario - due to the process of  photon-photon absorption in optically thick gamma-ray sources.

\section{Formation of sharp Klein-Nishina lines}
\label{KNlines}
\begin{figure}[t]
\includegraphics[width=0.45\textwidth]{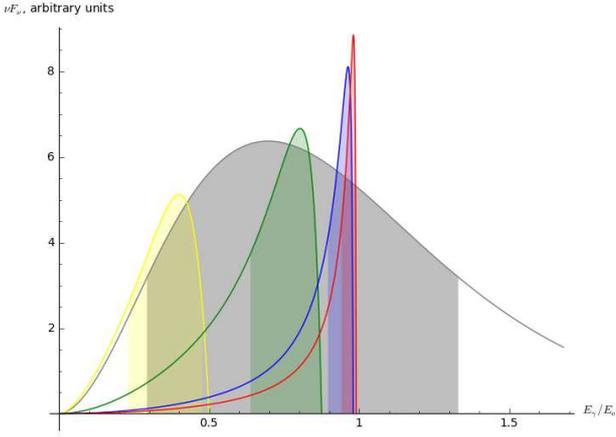}
\caption{{\it Colour}: energy spectra of the inverse Compton radiation of mono-energetic electrons upscattering isotropic target photons for 4 different values of the parameter $b$:  $1$, $7$, $50$ and $100$. 
The energy of gamma-rays is in units of the electron energy. 
{\it Grey}: the gamma-ray spectrum  produced by electrons  with  relativistic Maxwellian distribution;  in this case  the photon energies
are  in units of $4\Theta$, where $\Theta$ is the ``temperature'' of  Maxwellian distribution.}
\label{fig:spectra}
\end{figure}

\begin{figure}[t]
\includegraphics[width=0.50\textwidth]{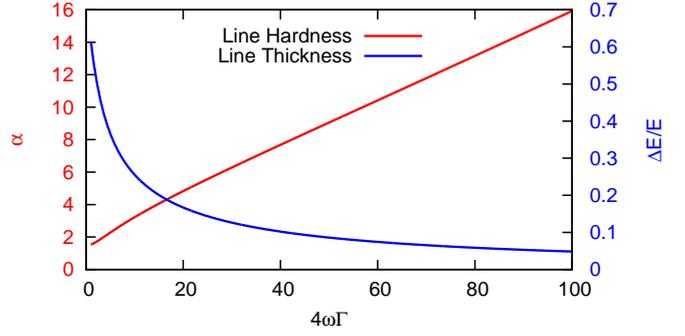}
\caption{The thickness and hardness of the Klein-Nishina line 
as functions of the $b$-parameter.}
\label{fig:line_prop}
\end{figure}

The strongest argument in favor of non-astrophysical origin of the $\sim 130$~GeV spectral 
feature is its very sharp profile.  It is very narrow, $\Delta E/E \approx {20{\rm\,GeV}\over{130\rm\,GeV}}\simeq15\%$,  with  an exponential rise and decay of the flux (which correspond  to
linear dependence as  seen in the semi-logarithmic scale plots), while 
in typical  astrophysical situations we expect much smoother spectral distributions. 
Nevertheless,  in the case of inverse Compton scattering  such profiles can be 
reproduced  by monoenergetic electrons provided that the scattering proceeds in deep Klein-Nishina regime. 
In this case the shape of the radiation spectrum is fully determined  by  a single  parameter,
 $b =4\omega\Gamma$, where  $\omega=\varepsilon/m_{\rm e}c^2$ and 
 $\Gamma=E_{\rm e}/m_{\rm e}c^2$ are the energy of the target 
 photon  in $m_{\rm e} c^2$ units and the electron Lorentz
factor, respectively.  In Fig. \ref{fig:spectra}  IC gamma-ray spectra are shown calculated for four 
different values of the parameter $b$: 1, 7, 50 and 100. 
The filled regions below each line correspond to the intervals where the flux level exceeds
50\% of the maxim value. Since the reported gamma-ray line flux  is  above the background only by a factor of two,  for characterization of the line profile we use the {\it half-height width} as a measure of the line thickness. This also allows us  to define the hardness of the left wing of the line by the power-law 
slope  at the point  where the flux
level reaches 50\% of the maximum value (the lower energy point). The line thickness and hardness determined in this way
are shown in Fig.~\ref{fig:line_prop} as functions of the parameter $b$.

In the Klein-Nishina  
regime,   the upscattered photons   repeat, to a large extent,   the spectrum of parent electrons. Therefore
even the Maxwellian distribution  of electrons, which is generally considered as a {\it very narrow}  one 
in the context of particle acceleration scenarios,   and which can be realized only with very specific conditions,  
cannot explain  the  reported very narrow line. This is illustrated in Fig.~\ref{fig:spectra}, where the 
IC gamma-ray spectrum produced by electrons with a Maxwellian distribution, ${\rm d}N_{\rm e}/{\rm d}\Gamma\propto\Gamma^2\exp{(-\Gamma/\Theta})$ is  shown;  the energy of gamma rays are expressed in units of $4\Theta$, where  $\Theta$ is the electron ``temperature''.  
It can be seen that the half-height width of this spectrum is
comparable to the energy at which the distribution maximum is achieved at  $E_\gamma \sim2.8\Theta$, and thus  much broader than the observed one.    Meanwhile,  the results 
presented in Fig.~\ref{fig:spectra}  show that the monoenergetic distribution of electrons does provide very sharp feature at $E_\gamma = E_{\rm e}$ with a width $\Delta E/E \leq   15 \%$, provided that the 
Klein-Nishina parameter $b \geq 30$. The corresponding hardness of the line, 
$\alpha =6$, also is in good agreement with observations. 
Note that the fixed value of the parameter $b$ 
implies a  monoenergetic distribution of target photons.  Although in specific  calculations one  
should assume more realistic  spectrum  of  seed photons,  this cannot significantly change the conclusions 
as long the parameter $b$ remains large.  
Given that for $b \gg 1$,  with a good accuracy $E_\gamma = E_{\rm e}$, 
one can immediately constrain the  energy range of target photons.  
\begin{equation}
\varepsilon \geq 15 (E_{\gamma}/130 \ \rm GeV)^{-1} \  \mbox{eV}\,,
\end{equation}
where $E_{\gamma}$ is the energy of the detected line.  
In the case of Planckian radiation field,  the above photon energy corresponds
approximately to  the radiation temperature of $\sim 5\times10^4$~K. 
In the case of isolated pulsars,  thermal emission from the surface of the neutron 
star would be  the main source of seed  photons for the formation of gamma-ray lines. 
The neutron star's surface temperature exceeds by two orders of magnitude this limit,
thus from  isolated pulsars we may expect extremely narrow gamma-ray lines.  In  binary systems with pulsars,   the seed photons are provided by companion stars with  radiation temperatures slightly less than the above estimate. Correspondingly, the gamma-ray lines would be broader 
with $\Delta E/E \geq 20 \%$; this still marginally agrees with observations of 
the 130 GeV line  from the galactic center region.  On the other hand, the efficiency of formation of Klein-Nishina lines from pulsar winds  in binary systems is significantly higher in binary systems than in isolated pulsars.

\section{Structure of gamma-ray line `hotspots'}
\label{Data}
An astrophysical origin of the 130~GeV line in general, and, more specifically,
its  association with pulsars, would  imply that the extended regions of excess emission  
are not truly diffuse structures, but rather represent superpositions of 
multiple unresolved  discrete gamma-ray sources. 
In order to  verify  this hypothesis we analyzed  52 month data  of \fermi/LAT using the latest \fermi software package v9r27p1.
The LAT aboard the \fermi satellite is a
pair-conversion gamma-ray detector operating between 20 MeV and 300 GeV. The \fermi
LAT has a wide field of view of $\sim 2.4$~sr at 1 GeV, and observes the
entire sky every two orbits. The details on  this  instrument can be found  in \citet{Atwood09}.


 \begin{figure}[t]
\includegraphics[width=0.45\textwidth]{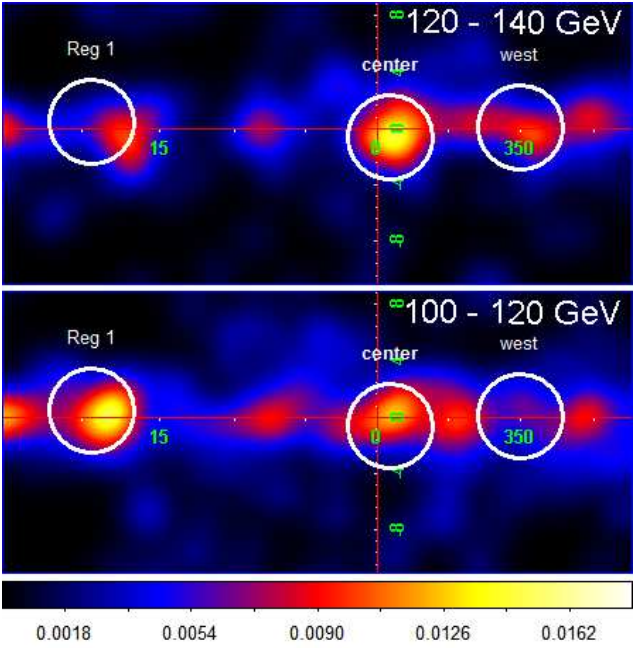}
\caption{The count maps  of  the ``Central'', ``West'' and $Reg~1$ regions 
(see the text)  in two energy intervals, 100-120 TeV and 120-140~TeV, 
smoothed with $3^\circ$ FWHM gaussian kernel}
\label{fig:regions_smooth}
\end{figure}

\begin{figure*}[t]
\includegraphics[width=0.29\textwidth]{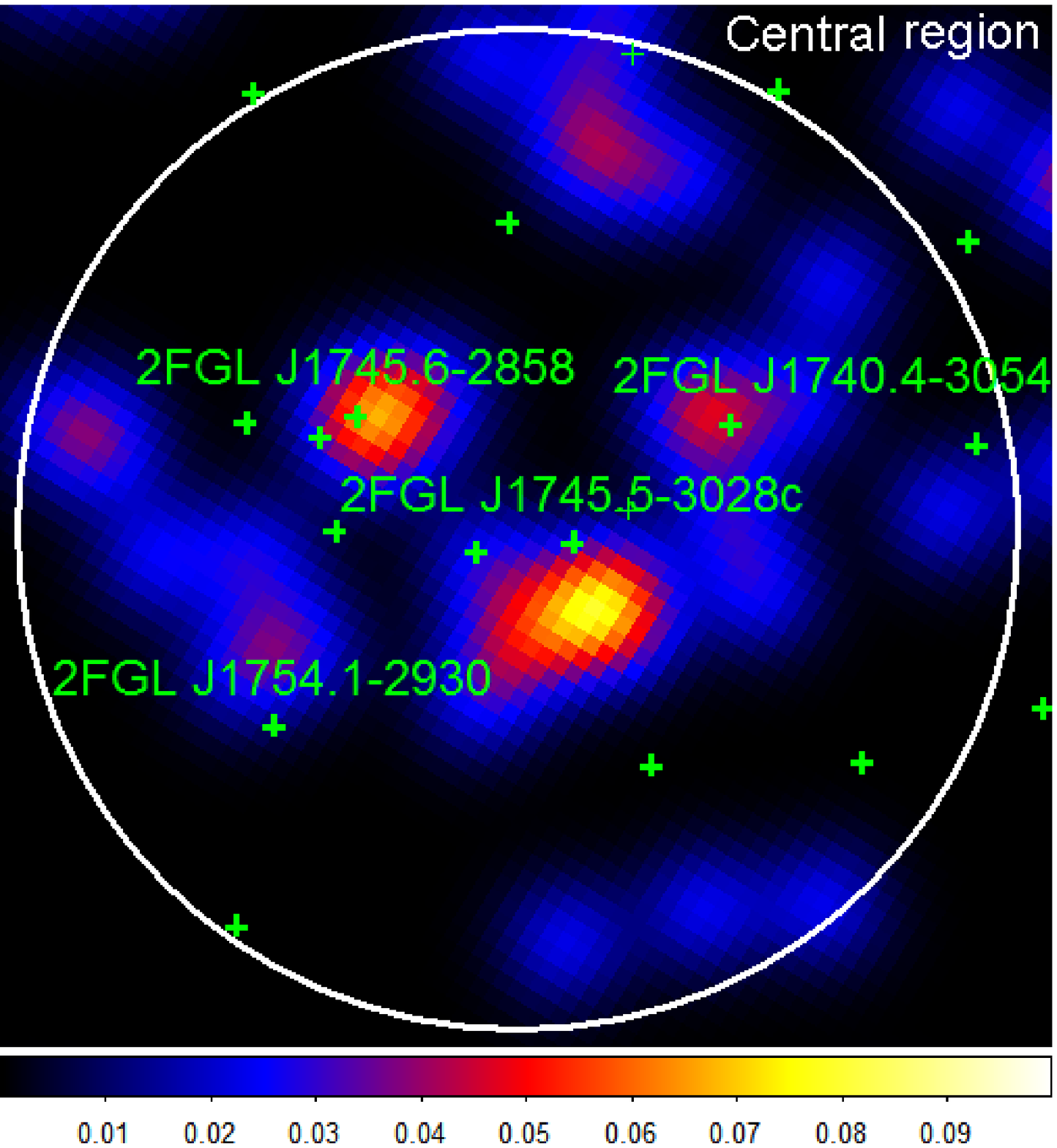}
\includegraphics[width=0.32\textwidth]{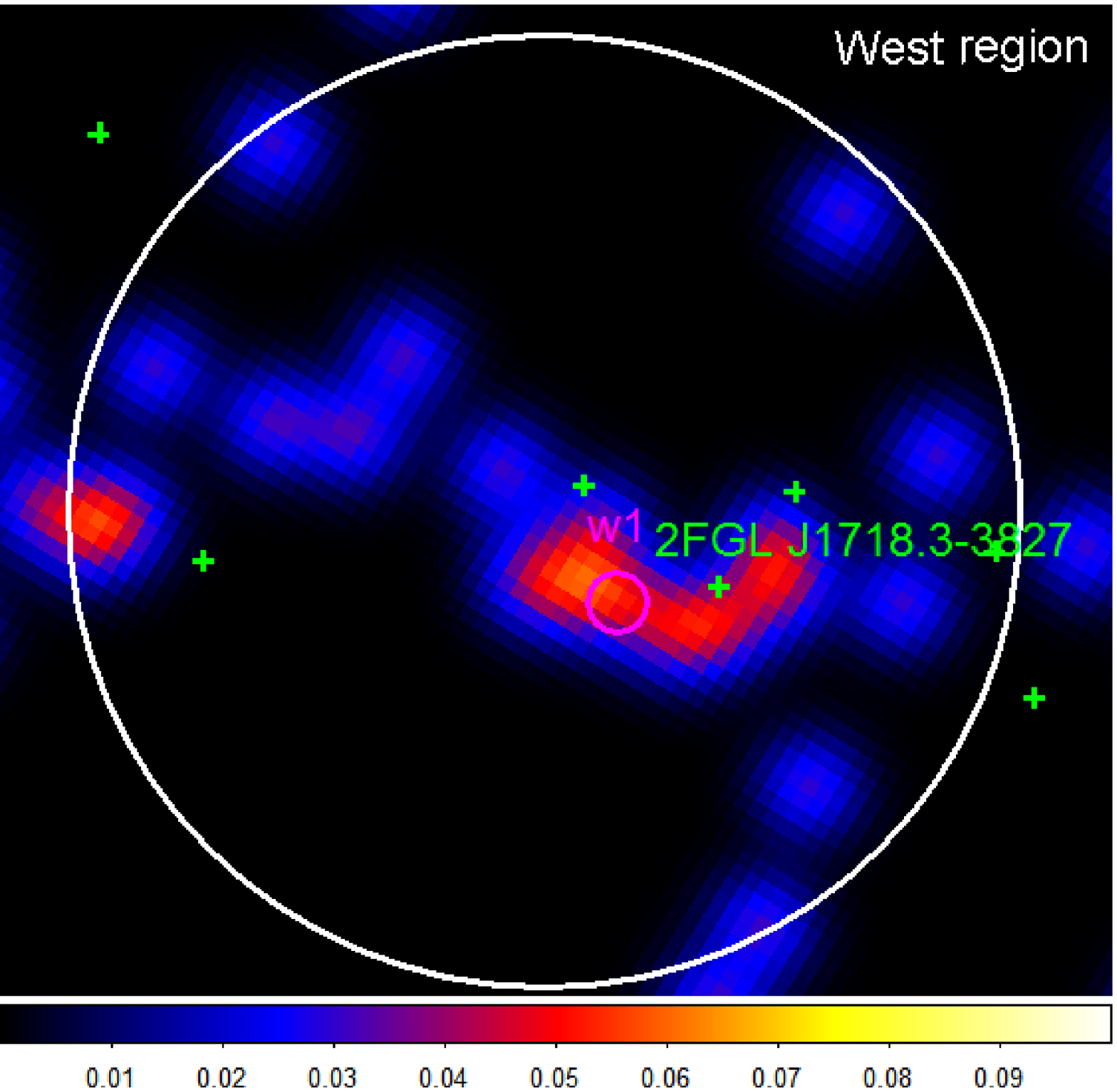}
\includegraphics[width=0.31\textwidth]{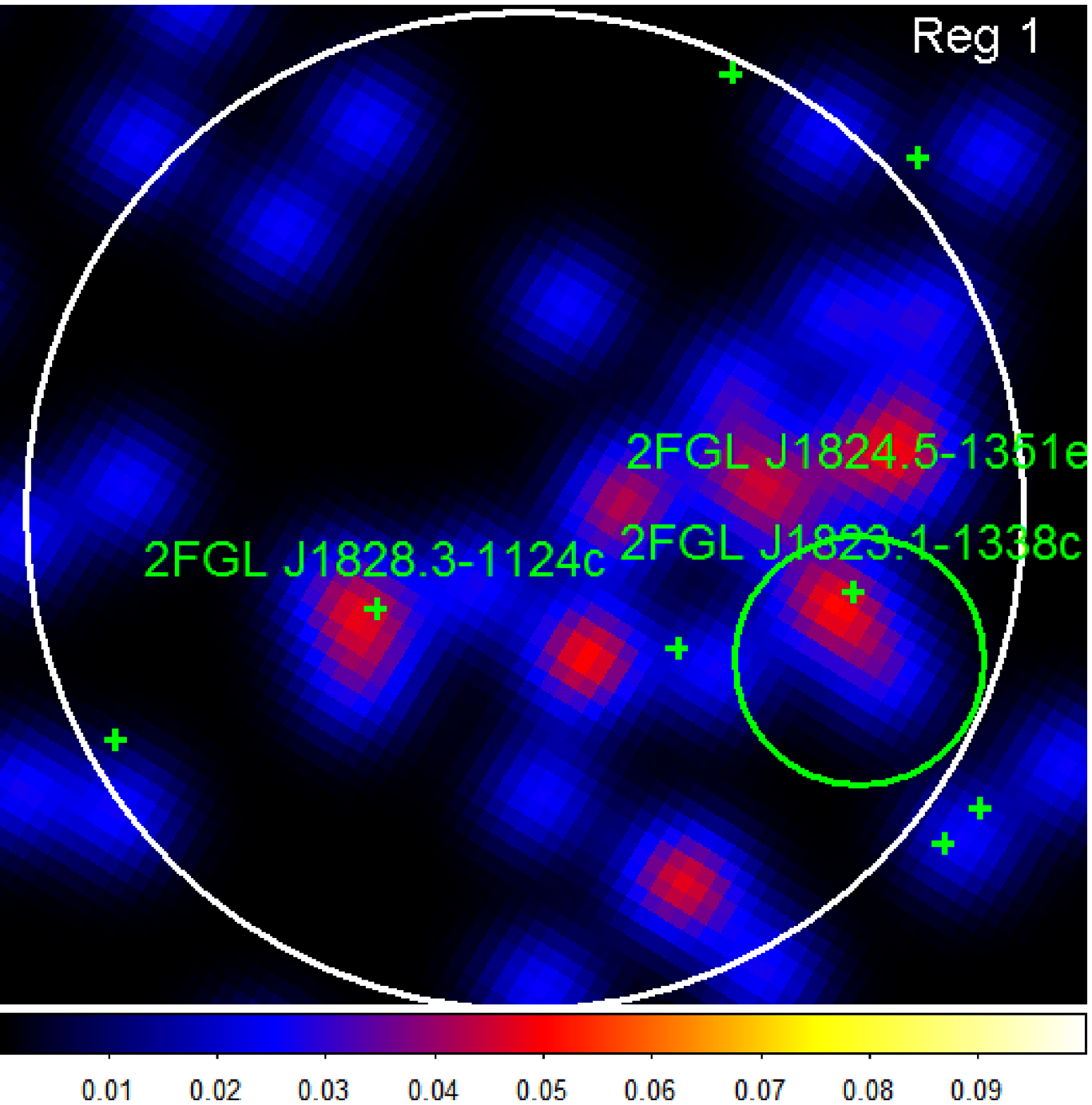}
\caption{Photon count maps of the ``Central'' (left) and ``West'' (middle) regions at 120-140~GeV  and the ``Reg~1'' region at 100-120~GeV energy bands smoothed with $0.25^\circ$ gaussian kernel ($\approx 95$\% \fermi PSF at 100~GeV). The corresponding regions from \citet{tempel12,boyarsky12} are shown with white circles, while the sources from the \fermi LAT two-year catalog are shown with green crosses. 
}
\label{fig:detailed_regions}
\end{figure*}

Here we consider three regions  from which  excess gamma-ray lines 
have been reported \citep{tempel12,su12,boyarsky12} in 
the energy interval  between 100-140~GeV. 
The  results  are presented  in Figs.~\ref{fig:regions_smooth} and \ref{fig:detailed_regions}, 
and summarized in Table~\ref{tab:reg_description}. 
\begin{table}
\begin{tabular}{|c|c|c|c|c|c|}
  \hline
  Region & Hot spot & $l$ & $b$ & R, deg & TS \\
  \hline
  Central &   & 359.0 & -0.7 & 3 &  \\
   &  2FGL J1745.6-2858  & 359.97 & -0.04 & 0.02 & 8.5 \\
   &  2FGL J1740.4-3054c & 357.73 & -0.08 & 0.16 & 8.4 \\
   & c1 & 359.02 & -1.41 & 0.08 & 7.5 \\
   & c2& 358.45 & -1.08 & 0.12 & 9.2\\
   & c3 & 358.43 &  1.47 & 0.23 & 8.6\\
  \hline
  West &  &350.0 &0.0 & 3 &  \\
   &w1  & 349.55 & -0.57 & 0.19 & 6.9 \\
  \hline
  Reg~1 & &19.38 & 0.40 & 3 &  \\
  \hline
\end{tabular}
\caption{The coordinates and names of considered regions and point-like hot spots detected in these regions. $l$ and $b$ are galactic coordinates of the region (new source), R -- the radius of the region (the radius of error-circle of the source). TS is the test-statistical value of the point-like hot spot (if $>5$). The significance of the hot spot can be estimated as $\sqrt{TS}\sigma$, see e.g. \citet{mattox96}}
\label{tab:reg_description}
\end{table}

In  Fig.~\ref{fig:regions_smooth} we show the count maps smoothed with 3$^\circ$ FWHM Gaussian 
kernel in  two energy intervals,   100-120~GeV and 120-140~GeV.  They are 
consistent with the previously reported  results.
The  locations  of  the ``Central'' and ``West'' regions  with excess emission at 120-140~GeV 
\citep{tempel12} and the ``Reg~1'' region with the excess emission  at 100-120~GeV 
\citep{boyarsky12},  are shown with white circles.  
 
In order to explore  the  spatial structures  of these regions we produced  
corresponding count maps smoothed with 0.25$^\circ$  gaussian kernel (comparable to the  \fermi PSF at 100~GeV). The corresponding maps are shown in Fig.~\ref{fig:detailed_regions}. The regions from \citet{tempel12} and \citet{boyarsky12} are shown with white circles. The green crosses show  positions of 
some of the GeV  gamma-ray sources from the
\fermi two-year source catalog.

In the ``Central''  region,  one can see several  ``hot spots''  (see Table~\ref{tab:reg_description} for the coordinates  and the test statistical (TS) values of  these  ``hot spots'' with TS $\geq 5$),  however the photon statistics in each of them is limited. 
The number of photons in five ``hot spots'' is 16 (from 32 total photons in the region), and the residual flux after removing the excesses is  comparable with the background 
(see e.g. \citet{boyarsky12}).  Thus, the  tendency of clumped  distribution of photons inside the ``Central'' region can be taken only as a hint  for presence of discrete sources of VHE gamma-rays.  It is interesting to note that the chance of false-positive detection  for the  two brightest  clumps (7 photons in total)  
over the  uniform photon distribution is  about 7\%.  Their possible association with two  hypothetical 
pulsars  seems a  quite  intriguing 
option in the context  of  interpretation of the possible two-peak structure of the 130~GeV line \citep{su12}  
as a superposition of Klein-Nisinia lines from two 
pulsar winds with different Lorentz factors.  The search for  candidate pulsars at these positions would be a straightforward  test  for verification of this 
hypothesis, although the detection of pulsars through their magnetospheric emission  
cannot be guaranteed  because of  possible misorientations of their  radiation cones.

The ``West'' (at 120-140~GeV) and ``Reg~1'' (at 100-120~GeV) regions  looks 
more diffuse in gamma-rays (see Fig.~\ref{fig:detailed_regions}, middle and right  panels),  however the poor 
photon statistics  do not allow any conclusion  concerning the level of clumpiness  in these regions.
While for  confirmation of gamma-ray line signals from these regions an increase of photon statistics 
by a factor of two or three would be adequate, the study of  spatial distributions of gamma-rays 
inside these regions  is a more demanding task and requires much higher photon statistics.

\section{Summary}
\label{summary}
In this paper we argue that there is a viable alternative to the 
DM origin of  the 130~GeV gamma-ray line as recently reported to be present in the 
galactic gamma-ray emission. We  demonstrate that  very sharp gamma-ray
spectral lines  can be produced by  pulsar winds through their Comptonizatioin,
predominantly by energetic (UV, X-ray)
radiation with  a relatively narrow  spectral distribution, thus the IC scattering
proceeds in the deep Klein-Nishina limit. In principle,
these conditions can  be fulfilled  both in isolated pulsars
and  binary systems.  The current  paradigm which connects
pulsars with their synchrotron nebulae through cold
ultrarelativistic electron-positron winds,  postulates
that the  electron-positron wind with a Lorentz factor between $10^4$ to $10^6$, carries
away almost the entire rotational  energy lost by the  pulsar. Thus, 
under the  condition of effective Comptanization of the wind,
a substantial fraction 
of the spin-down luminosity of a pulsar $L_{\rm rot}$
can be released   in a single gamma-ray line.   
 Depending on the intensity of illumination of the pulsar wind by surrounding radiation field(s), the efficiency of formation of such lines can be very high, formally close to 100 \%. Interestingly, the narrow profile  
of the 130~GeV line argues against such an extreme efficiency which can be realized in the case of 
optically thick source; this would imply not only strong  attenuation of gamma-rays  due to
photon-photon pair production, but also  significant broadening of the line because of the   
cooling of  electrons. However, both effects become  negligible in the case of  still very high, 
$\leq 10-20  \% $ efficiency of the wind Comptonization.  
Conservative estimates show that while such an efficiency 
can be readily achieved in binary systems, in the case of isolated pulsars 
the efficiency  is expected to be very low  unless the acceleration of the wind takes place close to
the light cylinder.  The  efficiency of transformation of  kinetic energy of the 
pulsar wind into Klein-Nishina gamma-ray  lines is a   key issue the discussion of which  is 
beyond the scope of this paper.  Clearly,  under certain conditions,  the efficiency can be 
quite high, and, for pulsars with spin-down luminosities exceeding 
$10^{36}$~erg/s and wind Lorentz factor $\geq 2 \times 10^5$,  
one may expect $\geq 100$~GeV gamma-ray lines 
with luminosities   exceeding  the bolometric luminosities 
of magnetospheric emission of pulsars. Moreover, one cannot exclude other configurations of 
compact objects, e.g. magnetars,  with (hypothetical)  relativistic electron-positron winds, as effective 
multi-GeV gamma-ray emitters.

In this regard
a natural question arises:
if  the pulsars can  work as  effective generators of  VHE gamma-ray lines,
why have they not been detected yet?  There could be several answers to this question.
In particular, in many objects the  target radiation field could not be sufficient
for effective conversion of the kinetic energy of the wind to IC gamma-radiation.
Alternatively, if the wind Lorentz factor is small and/or the target photons have a broad distribution, the IC scattering of electrons in the Thompson regime would lead to  a gamma-ray continuum  the detection and identification of which could appear not an easy task.
The  formation of the line  is  effective only for pulsars with wind Lorentz factor
exceeding $10^5$; the corresponding gamma-ray line is formed around or above 100~GeV.
At these energies the potential of \fermi LAT is limited because of the small detection area.
On the other hand, the current  imaging Cherenkov telescopes operate 
effectively above 100~GeV, so could simply have missed the  lines around 100~GeV.  
Over the next several years  \fermi LAT  can  double the photon statistics which will be 
sufficient,  hopefully, 
for a  firm detection  of the 130~GeV line, but still not adequate for morphological studies of the 
gamma-ray line emitting regions. In this regard more promising 
seem to be  forthcoming studies with  new, low-energy threshold imaging
 atmospheric Cherenkov telescope systems, 
in particular by the new 27~m diameter dish of the H.E.S.S.  telescope array which is  located 
in a perfect site in the Southern Hemisphere for observations of the galactic center region.  This new 
instrument with energy threshold as low as 50 GeV,  huge collection area exceeding $10^4 \ \rm m^2$,
and energy resolution close to 20 \%   should allow  (in the near future!) deep spectroscopic and morphological studies of the inner galaxy in multi-GeV gamma-ray lines.   If confirmed, the existence of such lines may lead to an exciting new research area --  {\it Klein-Nishina  gamma-ray line astronomy} -- that will open the way for future ground-based gamma-ray detectors, in particular the Cherenkov Telescope Array (see   http://www.cta-observatory.org/) to probe the physics and astrophysics of  relativistic outflows, in particular pulsar winds. 

\begin{acknowledgements}
We  gratefully acknowledge useful conversations  with Segey Bogovalov and Roland Crocker. 
\end{acknowledgements}


\end{document}